\documentstyle[aps]{revtex}

\begin{document}
\draft


\title{Description of Double Giant Dipole Resonance within
the Phonon Damping Model}

\author{
C.~A.~Bertulani$^{1}$, P.~F.~Bortignon$^{2}$, V.~Yu.~Ponomarev$^{3,4}$,
and V.~V.~Voronov$^{3}$}
\address{
$^{1}$Brookhaven National Laboratory, Physics Department,
Nuclear Theory Group, Upton, NY 11973-5000, USA
}
\address{
$^{2}$Dipartimento di Fisica, Universit\"a di Milano and
INFN Sez. di Milano,
Via Celoria 16,
I-20133 Milano, Italy
}
\address{
$^{3}$Bogoliubov Laboratory of Theoretical Physics,
Joint Institute for Nuclear Research,
Dubna, Moscow region, Russia
}
\address{
$^{4}$Institut f\"ur Kernphysik, Technische Universit\"at Darmstadt, 
D--64289  Darmstadt, Germany
}

\date{\today }
\maketitle


~

In a recent Letter~\cite{d1}, an overall agreement with the experimental
data for the excitation of the giant dipole resonance (GDR) and double
giant dipole resonance (DGDR) in relativistic heavy ion collision (RHIC)
in $^{136}$Xe and $^{208}$Pb nuclei has been reported. The phenomenological
Phonon Damping Model (PDM) has been used. The strong (about a factor two)
enhancement of the DGDR cross section in $^{136}$Xe \cite{Sch93}
apparently reproduced in the Letter, has been a challenge for many years
and in spite of several attempts (see, e.g. \cite{Aum98,Bert99}), it
remains open. In this Comment, we point out that the agreement with the
experimental findings in Ref.~\cite{d1} is achieved by a wrong calculation
of the DGDR excitation mechanism.

To calculate the DGDR cross section, an expression (Eq.~(6)) is used
which is similar to the photoabsorption cross section with
the GDR strength function replaced by the one of the DGDR.
Then, it is inserted in Eq.~(7) which thus describes the DGDR excitation
in a direct or one-step process \cite{note}.

An essential difference between nuclear reactions with real and
virtual (as in RHIC) photons is that multi-step processes with the
sequential absorption of two (three, etc.) virtual photons may take
place in the last reaction. Theoretically, it is described, e.g. in the
second (third, etc.) order perturbation theory \cite{Win79,Ber88}.
The reduction of the two-step process to the one-step one is
not possible on the physical grounds.

Whether the DGDR is excited in RHIC in one- or two-step process is
presently not in question.
Firstly, microscopic calculations with no free parameters indicate that
the two-step process is stronger by at least two-orders of magnitude or
even more \cite{Bert99}. Secondly, it has been proved experimentally
\cite{Bore96}.

Eq.~(6) contains a scaling factor $c^{(2)}$ which determines the absolute
value of the DGDR cross section. It is computed by
equating first order (Eq.~(7)) and second order (Eq.~(8)) expressions.
To equalize the values, known to be different by orders of magnitude
(see, above), appears to us not correct.

In addition, we find it difficult to call ``microscopic calculations''
the fits performed in the frame of
the PDM to obtain strength functions, as done in \cite{d1}, when the
strength parameter $F_{ph}$ changes by two orders of magnitude when
contributions of higher-order processes to the damping of the GDR are
included, as in the extension of the model in Ref. \cite{d2}. This
means that the main mechanism for the GDR and DGDR widths is missed in
Ref.~\cite{d1} and that the agreement with the data is achieved by an
unrealistically large value of $F_{ph}$.
In Ref. \cite{d2} they claim that the higher-order processes may be
effectively accounted for by renormalization of $F_{ph}$.
This is not correct because the diagrams of high-order graphs (see,
Figs. 1b-e in Ref.~\cite{d2}) cannot be reduced to the diagram of
the lowest-order one (Figs.~1a) with a renormalized vertex.
Again, as for the cross-section above, their ``effective'' treatment
contradicts the ``right physical content''.

The PDM fits yield the value of $F_{ph}$ in the heavier double-magic
$^{208}$Pb larger than in the lighter semi-magic $^{136}$Xe, while it
is clear from general arguments that it should be the opposite.

The parameter $F_{ph}$ also enters  in the calculation of the DGDR
anharmonicity, i.e. in the deviation of the DGDR energy centroid from
twice the energy value of the GDR. Thus, it becomes clear why the PDM yields
much larger ``anharmonicities'' than microscopic calculations \cite{Pon00}.

In our opinion, any conclusion drawn from those calculations about the
deviation of the DGDR properties from the harmonic limit expectations appears
rather difficult to understand.



\begin{references}
\bibitem{note}
The DGDR has in fact components with quantum numbers $0^+$ and $2^+$.
Thus, to calculate the photoexcitation of the DGDR (a process never been
observed because of the extremely small cross section \cite{Bert99}),
the cross section should be cubic on the excitation energy and not linear
as in Eq.~(6).
\end{references}
\end{document}